\documentstyle[preprint,aps]{revtex}
\tighten
\begin{document}
\widetext
\input epsf
\preprint{NYU-TH/98/10/04, HUTP-98/A065, NUB 3187}
\bigskip
\vspace{0.2in}

\title{BPS Domain Walls in Large $N$ Supersymmetric QCD}
\medskip
\author{Gia Dvali$^{1,2}$\footnote{E-mail: gd23@scires.nyu.edu}, 
Gregory Gabadadze$^{1}$\footnote{E-mail: 
gg32@is9.nyu.edu} and 
Zurab Kakushadze$^{3,4}$\footnote{E-mail: 
zurab@string.harvard.edu}}

\bigskip
\address{$^1$Department of Physics, New York University, 4 Washington Place,
New York, NY 10003\\ 
$^2$International Centre for Theoretical Physics, 34100 Trieste, Italy\\
$^3$Lyman Laboratory of Physics, Harvard University,
Cambridge,  MA 02138\\
$^4$Department of Physics, Northeastern University, Boston, MA 02115}
\date{January 10, 1999}
\bigskip
\medskip
\maketitle

\begin{abstract}
{}We explicitly construct BPS domain walls interpolating between neighboring
chirally asymmetric vacua in a model for large $N$ pure supersymmetric QCD.
The BPS equations for the corresponding ${\bf Z}_N$ symmetric
order parameter effective Lagrangian reduce to those in the $A_N$ 
Landau-Ginsburg model assuming that the higher derivative terms 
in the K{\"a}hler potential are suppressed in the large $N$ limit. 
These BPS domain walls, which have vanishing
width in the large $N$ limit, can be viewed as 
supermembranes embedded in a (3+1)-dimensional supersymmetric 
QCD background. The supermembrane couples to a three-form 
supermultiplet whose components we identify with the 
composite fields of supersymmetric QCD. We also discuss certain aspects of 
chromoelectric flux tubes (open strings) ending on these walls   
which appear to support their interpretation as D-branes. 
\end{abstract}

\section{Introduction}

{}Supersymmetric gauge theories provide opportunities 
to gain analytic control over certain complicated non-perturbative 
phenomena in (3+1)-dimensional strongly coupled gauge models
\cite {wittenind,beta,affleck,SV,am,seiberg,switten,intriligator}.  
The closest supersymmetric 
cousin of QCD, ${\cal N}=1$ supersymmetric $SU(N)$ gluodynamics (pure
supersymmetric QCD), is believed to be a confining gauge theory with a 
mass gap and chiral symmetry breaking \cite {wittenind,SV,am}.
One can therefore expect that a better understanding of pure supersymmetric 
QCD (SQCD)\footnote{Throughout this paper
``SQCD'' refers to the ${\cal N}=1$ 
supersymmetric $SU(N)$ gauge theory without matter, {\em i.e.}, pure 
supersymmetric gluodynamics.} 
might shed some light on non-perturbative phenomena in conventional Quantum
Chromodynamics.

{}Chiral symmetry breaking in SQCD is due to 
gluino condensation. The composite operator 
${\rm Tr}(\lambda^{\alpha} \lambda_{\alpha})\equiv \lambda\lambda$ 
acquires a non-zero vacuum expectation value (VEV) \cite{wittenind,SV}: 
$\langle\lambda\lambda\rangle_k= N
\exp(2\pi ik/N) \Lambda^3$, where $\Lambda$ is 
the dynamically generated scale of SQCD. 
Thus, there are $N$ inequivalent vacua related by ${\bf Z}_N$ 
discrete symmetry  and labeled by the phase of the gluino 
condensate $k=0,\dots,N-1$. In each particular vacuum state
the discrete ${\bf Z}_N$ symmetry is broken. As a result, 
there should exist domain walls separating pairs of distinct vacua. 

{}In \cite{wall} it was shown that the ${\cal N}=1$ supersymmetry (SUSY) 
algebra admits a nontrivial central extension if BPS 
saturated domain walls are present in the theory. Assuming that such BPS walls
exist, their tension can be calculated exactly on the basis 
of the SUSY algebra alone \cite{wall}. Whether 
the BPS saturated domain walls are indeed present in a given
model is a dynamical issue.
The strong coupling nature of SQCD, however, makes it difficult to study
this problem in terms of colored fields. In order to circumvent this
difficulty, one can use  an  order parameter effective Lagrangian
relevant for the  chiral symmetry breaking.  Thus, in the 
large $N$ SQCD  one could attempt to  solve the corresponding 
BPS equations in terms of colorless composite 
variables (order parameters). 
This approach, as we will discuss later, 
can be argued to be reliable in the limit of infinitely large $N$. 
This is the route we are going to follow in the present work.
In particular, we will explicitly construct BPS saturated domain walls 
interpolating between neighboring chirally asymmetric vacua 
in a model for large $N$ SQCD. The width of these walls is vanishing in the large
$N$ limit. We argue that this is precisely the property which 
gives rise to the wall tension consistent with the prediction of the SUSY 
algebra \cite{wall}. 

{}Having found the BPS walls in the large $N$ limit, we 
discuss more general questions related to physics of extended objects
in SQCD. In fact, we address the issue whether the lowest-spin
states of SQCD can be viewed as excitations of such 
extended objects. Our discussion is motivated by the relation 
of the large $N$ gauge models and string theories. Indeed, 
in \cite{thooft} it was shown that in the large $N$ limit 
the gauge theory diagrams are organized in terms of Riemann
surfaces with boundaries and handles, and addition 
of each extra handle on the surface corresponds to 
suppression by a factor of $1/N^2$. This observation leads to the hope that 
the large $N$ 
gauge theory might be described by some kind of string theory. Then the 
large $N$ expansion of gauge theories would be mapped to the string
expansion in terms of properly weighted world-sheets of 
various topologies.

{}The first concrete realization of this idea in the context of 
weakly coupled gauge theories 
({\em i.e.}, when the effective
gauge coupling $\lambda=Ng^2_{YM}$ is fixed at a value 
$\lambda{\ \lower-1.2pt\vbox{\hbox{\rlap{$<$}\lower5pt
\vbox{\hbox{$\sim$}}}}\ }1$) was given in \cite{CS} in the framework of
three-dimensional Chern-Simons gauge theory where the boundaries of the
string world-sheet are ``topological'' 
D-branes\footnote{For recent developments in these directions, see 
\cite{GV}.}. More recently, in
\cite{BKV}\footnote{This was subsequently generalized 
to include unoriented world-sheets in \cite{ZK}.} 
string expansion was shown to precisely 
reproduce 't Hooft's large $N$ expansion for certain four dimensional 
(super)conformal gauge theories. 

{}In strongly coupled gauge theories, however, the story 
appears to be much more involved. In particular, it is 
unknown what is the string theory that governs the dynamics of
the QCD string which is expected to arise as an effective 
description in the large $N$ limit. This string theory is 
expected to be non-critical \cite{polyakov}, which makes it difficult 
to study. Nonetheless, in \cite{witten} Witten has made an interesting
observation that there might be a connection between SQCD 
strings and domain walls in the large $N$ limit. In particular, based
on the assumption that the SQCD domain walls are BPS saturated \cite{wall}, 
Witten argued that in the large $N$ limit the domain walls 
connecting two vacua labeled by $k$ and $k^\prime=k+1$ appear 
to be objects that look like
D-branes \cite{Pol} on which the SQCD string should be able to end. 
Such D-brane-like domain walls have also appeared in 
\cite{witt} in the context of large $N$ non-supersymmetric QCD. 

{}We argue in this work that
in the large $N$ limit the BPS domain walls with the vanishing width can be
viewed as supermembranes embedded in an SQCD background.
Moreover, the component fields of the three-form supermultiplet 
which couples to the supermembrane can be identified in 
terms of the composite fields of
SQCD. We also point out that some of the lowest-spin 
states of SQCD can be viewed as excitations of a closed string which propagates
in the vacua separated by the supermembranes. 
A BPS supermembrane can be regarded as a
D-brane in the large $N$ limit if there are open strings ending on it.
We discuss certain aspects of chromoelectric flux tubes ending on the
domain walls \cite{flux} which appear to support the D-brane interpretation 
of these objects.   

{}The rest of this paper is organized as follows. In section II we 
discuss the large $N$ behavior of a generic SQCD configuration and 
point out why a BPS state can exist 
in the large $N$ limit. Then, we briefly discuss the ${\bf Z}_N$
symmetric order parameter effective Lagrangian for SQCD
(see \cite {VY,KovShif} and \cite {gabad}),
and also give a general discussion of domain walls in ${\cal N}=1$ 
supersymmetric theories. In section III we explicitly 
construct BPS domain walls using
a model ${\bf Z}_N$ symmetric order parameter effective superpotential
\cite {gabad}, 
by solving the corresponding BPS equations
in the large $N$ limit. 
Note that large $N$ BPS domain walls
have been constructed in \cite{gia}. The idea of \cite{gia} is to
avoid certain complications arising in SQCD by taking an indirect route. 
In \cite{gia} it was argued that in the large $N$ limit 
the problem effectively reduces to the question of finding  BPS solitons in  
the large $N$ $A_N$ Landau-Ginsburg theory. 
In this paper we construct the BPS domain walls {\em directly} 
within the framework of the ${\bf Z}_N$ symmetric 
model of  large $N$ SQCD. 
We show that the corresponding 
system of BPS equations indeed reduces to that in  
the large $N$ $A_N$ Landau-Ginsburg model provided that higher derivative
terms in the K{\"a}hler potential are suppressed in the large $N$ limit.  
This way we {\em derive} the results  of \cite{gia} which were obtained 
via an indirect construction. In section IV 
we elucidate some relations to other works on the subject.
In section V we present our discussions and conclusions. 

\section{${\bf Z}_N$ Symmetric Vacua and Domain Walls}\label{prelim}

{}In this section we review some facts about the large $N$ limit of
${\cal N}=1$ supersymmetric QCD. We argue that the SQCD 
domain wall configurations can saturate the BPS bound
if the width of the walls vanishes in the large $N$ limit.   
We will also discuss the ${\bf Z}_N$ symmetric order parameter effective 
superpotential of  \cite {gabad} which effectively encodes 
dynamics relevant for describing large $N$ domain walls.

\subsection{Gaugino Condensation and Domain Walls in SQCD}

{}Consider ${\cal N}=1$ supersymmetric QCD with $SU(N)$ gauge group and no
matter. Let $\Lambda$ be the dynamically
generated scale of the model. In this theory the gluino condensate is given
by \cite{SV} \begin{equation}\label{cond}
 \langle \lambda\lambda\rangle _k= N \exp(2\pi i k/N) \Lambda^3~.
\end{equation}
Here $k=0,\dots,N-1$, that is, there 
are $N$ inequivalent vacua corresponding to $N$ different
phases for the gaugino condensate. 
Note that the overall factor of $N$ in (\ref{cond}) follows from the fact that
the gaugino condensate $\langle\lambda\lambda\rangle$ 
involves a trace over the gauge indices, and the resulting VEV
should contain the corresponding second Casimir factor.

{}According to  (\ref{cond}), 
there are $N$ inequivalent vacua with spontaneously broken discrete symmetry
in SQCD.  
Hence, there should exist domain walls separating  these different vacua 
\cite{wall}\footnote{A number of interesting properties of domain walls
in various models were discussed in 
\cite{rel1,rel2,rel22,rel23,rel24,rel25,rel26,smilga,rel3,rel31,kaplun,troitsky}.}. 
On dimensional grounds
we expect the tension of such a wall to be of order $\Lambda^3$. It is,
however,  possible to compute the tension exactly in terms  of the gluino
condensate provided that they are BPS saturated \cite{wall}. Consider the
domain wall separating the vacua with $\langle \lambda\lambda\rangle_k=
N \exp(2\pi i k/N) \Lambda^3$ and
$\langle \lambda\lambda\rangle_{k^\prime}= N
 \exp(2\pi i k^\prime/N) \Lambda^3$.
The tension of the wall 
(provided that it is BPS saturated) is defined by the central
charge $Q_{kk^\prime}$ 
in the corresponding central extension of the ${\cal N}=1$ superalgebra. The
central charge is 
proportional to the absolute value of the difference between the values
of the superpotential in the two vacua:
\begin{equation}
 Q_{kk^\prime}~\propto ~\vert {\cal W}_k-{\cal W}_{k^\prime}\vert~.
\end{equation}
We therefore have the following tension for the BPS domain walls:
\begin{equation}\label{tension}
 T_{kk^\prime}={ N^2\over 4\pi^2}
 \left| \sin\left({\pi(k-k^\prime)\over N}\right) \right|
 \Lambda^3~.
\end{equation}

{}It is, however, difficult to explicitly construct these domain walls 
(or even check that they are BPS saturated) in SQCD: the order 
parameter $\lambda\lambda$ is a composite operator, and we are dealing 
with a strong coupling regime. 
Instead, we will be  able to explicitly find 
BPS domain walls in {\em large} $N$ SQCD
using the ${\bf Z}_N$ symmetric order parameter 
effective superpotential \cite {gabad}.

{}The presence of BPS saturated domain walls in large $N$
SQCD has important implications \cite{witten}. 
Strongly coupled SQCD in 
the large $N$ limit is believed to 
be described by a string theory with 
the string coupling $\lambda_s\sim 1/N$.
Extended solitons in this 
string theory are expected to have tension which 
goes as $1/\lambda^2_s\sim N^2$. Let us, however, consider BPS domain walls 
with $k^\prime=k+1$. In the large $N$ 
limit their tension goes as $\sim N\sim 1/\lambda_s$. 
In \cite{witten} it was suggested that such domain walls can be 
viewed as D-branes \cite{Pol} rather than solitons  in the SQCD string
context.  Open SQCD strings then can end on these D-branes \cite{witten}.   
    
\subsection{Domain Walls at Large $N$}

{}Let us consider the domain walls with $k^\prime=k+1$ in the large $N$ limit.
According to (\ref{tension}) the tension of such a domain wall 
(provided that it is BPS saturated)
is given by
\begin{eqnarray}\label{tension1}
 T_D = {N\over 4\pi}~\Lambda^3.
\label{BPSenergy}
\end{eqnarray}
The fact that $T_D$ scales as $N$ in the large $N$ limit might appear a 
bit puzzling
if one thinks of such BPS domain walls as SQCD solitons. Indeed, the number of
degrees of freedom (gluons and gluinos) in SQCD scales as $N^2$ in the large
$N$ limit, so naively one expects the energy density of a generic SQCD
configuration to scale not as $N$ but as $N^2$.  If this argument were
precise, it would imply that the large $N$ SQCD domain walls with
$k^\prime=k+1$ could not be BPS saturated. However, in the following we will
argue that  the $N$ dependence of the energy density $T_D$ given by
(\ref{tension1}) is indeed correct for the corresponding BPS saturated SQCD
solitons\footnote{We are grateful to M. Shifman for useful conversations on
the issues discussed in this subsection.}.  

{}In the large $N$ limit the effective Lagrangian is expected to have 
the following form:
\begin{eqnarray}\label{G}
{\cal L} = N^2 ~{\cal L}_0 (\Phi, \nabla \Phi, \nabla ^2 \Phi, \dots)~, 
\end{eqnarray} 
where $\Phi$ collectively denotes composite colorless bound states of 
gluons and/or gluinos. 
The corresponding fields $\Phi$ are rescaled in such a way that 
all the leading $N$ dependence in (\ref{G}) is in the overall factor of $N^2$.
In particular, the equations of motions for $\Phi$ do not contain any factors
of $N$.  Thus, in the large $N$ limit the volume energy density of any
solitonic configuration $\epsilon\sim N^2 \Lambda^4$. This implies that the
tension of a domain wall $T\sim\epsilon d$ is proportional to $N^2\Lambda^3$
if the
``width'' of the domain wall $d\sim 1/\Lambda$. However, if  $d\sim
1/N\Lambda$, then we have $T\sim N\Lambda^3$. This indicates that the domain
walls with $k^\prime=k+1$ should have vanishingly small width
$d\sim1/N\Lambda$ in the large $N$ limit \cite{gia}. In section III we will
explicitly show that this is indeed the case. Here, however, we will give a
simple explanation of why $d\sim1/N\Lambda$ for such domain walls.   

{}Since all the fields $\Phi$ in ${\cal L}_0$ are such that $|\Phi|\sim 1$ in the large $N$ limit, 
solitonic solutions with $d$ depending on $N$ might at first seem impossible to accommodate.
Thus, naively one might expect that ${\cal L}_0$ has no ``knowledge'' of $N$. This, 
however, is {\em not} the case provided that the number of inequivalent vacua 
described by ${\cal L}_0$ scales as $N$ \cite {flux}, 
which is precisely what happens in SQCD. 
This way the Lagrangian ${\cal L}_0$ does ``know'' about $N$. 
As a simple toy 
example of such a system consider a single chiral superfield $\Phi$ with the scalar potential
${\cal V} (\Phi)=N^2|1-\Phi^N|^2$. There are $N$ inequivalent vacua in this theory: 
$\Phi_k=\exp(2\pi i k/N)$, $k=0,1,\dots, N-1$. Thus, even though
$|\Phi_k|\equiv 1$ in ${\cal V}$, we have $|\Phi_{k+1}-\Phi_k|\propto 1/N$
in the large $N$ limit.  

{}In the following we will show that the model effective Lagrangian
for supersymmetric gluodynamics \cite {gabad} 
possesses precisely the properties discussed
above.  In fact, among other things it describes $N$ inequivalent
vacua with the broken chiral symmetry. In the large $N$ limit
it admits BPS solitons with the width proportional to $1/N$. 
These solitons are the BPS domain  walls separating 
neighboring chirally asymmetric vacua of the theory.

\subsection{The Order Parameter Effective Action}\label{effective}

{}The $N$ chirally asymmetric vacua in SQCD are defined by the 
VEV of the order parameter 
$\lambda\lambda$, the gluino  bilinear. To construct the domain walls 
interpolating between different vacua we can try to write down an effective Lagrangian
for this order parameter. The full effective Lagrangian (\ref{G}) would contain infinite 
number of fields corresponding to higher dimensional operators of the theory
and higher derivatives as well.
However, in the large $N$ limit we expect certain simplifications, in fact, the
higher dimensional operator contributions should be suppressed by 
extra powers of $N$. More
precisely, we expect this to be the case for domain walls interpolating
between neighboring vacua; in this case the relative change 
in the gluino  condensate
$\langle\lambda\lambda\rangle$ is ${\cal O}(1/N)$, so that we expect the
adequate description in the large $N$ limit to be in terms of a
truncated\footnote{By ``truncated'' we mean that the corresponding effective
action is obtained from the full effective action (\ref{G}) by keeping a
finite number of fields $\Phi$ and neglecting  the rest of the fields in the
large $N$ limit.} effective Lagrangian containing a finite number of fields. 
On the other hand, if we consider domain walls at finite $N$ the
relative change in the gaugino condensate $\langle\lambda\lambda\rangle$ is no
longer small, so that truncating the full effective action (\ref{G}) to a
finite number of lowest dimensional operators may no longer be justified. A
similar remark applies to the case of large $N$ domain walls with
$|k^\prime-k|\sim N$. In the following, therefore, we will focus on the large
$N$ limit where we will confine our attention to the domain walls with
$k^\prime=k+1$.    

{}Thus, we are going to be interested in deducing the effective Lagrangian adequate 
for studying large $N$ domain walls. It is reasonable to assume \cite{VY} 
that the order parameter effective Lagrangian should be constructed in terms
of the chiral superfield  \begin{eqnarray}
 S\equiv \langle {\mbox{Tr}} (W_{\alpha}W^{\alpha}) \rangle=
 \langle {\mbox{Tr}} (\lambda^{\alpha}\lambda_{\alpha}) \rangle+\dots\equiv
 \langle \lambda\lambda\rangle+\dots~,
\nonumber
\end{eqnarray}
where $S$ is regarded as a classical superfield, and the matrix elements 
are defined in the presence of an appropriate background source (super)field
(for detail see \cite {Shore}). Note
that the gluino bilinear $\lambda\lambda$ is the lowest component of the
chiral superfield ${\mbox{Tr}} (W_{\alpha}W^{\alpha})$. (Here $W_\alpha$ is
the usual gauge field strength chiral superfield.)  Then, the order parameter
effective superpotential  reproducing all the anomalies of the model is given
by \cite {VY}: \begin{eqnarray}\label{vy}
{\cal W}_{VY}=NS \left[\ln\left({N\Lambda^3 \over S}\right)+1\right]~.
\end{eqnarray}
The corresponding scalar potential  
describes spontaneous chiral symmetry breaking with a 
non-zero gaugino condensate
\cite {VY}. 

{}The above effective superpotential, however, is not adequate for studying 
domain walls in SQCD. This can be seen as follows. First, the superpotential
(\ref{vy}) does not respect the ${\bf Z}_N$ discrete symmetry \cite {KovShif}.
This, in particular, implies that the use of (\ref{vy}) cannot be justified
when describing domain walls interpolating between different chirally
asymmetric vacua of the model \cite{KovShif}. Thus, the superpotential
(\ref{vy}) is only adequate for describing a given chirally asymmetric vacuum.
In fact, as it stands in (\ref{vy}), such a description is only appropriate
for the vacuum with $k=0$. This is related to  the strong CP violation.
Let us notice that there is no CP violation in SQCD
since  the 
CP odd $\theta$ angle  can always be canceled by chiral transformations of
massless gluino fields. 
Let us go back to the effective Lagrangian 
and see how this property is realized there. 
As we mentioned above, the lowest component of the 
chiral superfield $S$ is the gluino bilinear $\langle \lambda
\lambda \rangle$. The highest component of the superfield 
$S$ is formed by the composite ``glueballs'': 
$\langle G_{\mu\nu}G^{\mu\nu}+iG_{\mu\nu}{\tilde G}^{\mu\nu} \rangle$.
As usual, these latter fields are auxiliary components  of 
the chiral superfield  $S$, and, therefore, do not 
appear in the on-shell scalar 
potential of the model\footnote{These auxiliary fields 
can appear in the effective Lagrangian if higher  superderivatives 
are present in the 
corresponding K{\"a}hler potential. In this case the F-terms become
dynamical fields. However, including these fields as dynamical degrees of 
freedom is accompanied by 
problems with positive definiteness of the corresponding scalar
potential (for details see \cite{Shore}), and, therefore, 
it is assumed that the K{\"a}hler potential does not contain
superderivatives. This is expected to be the case in the large $N$ limit}.
Thus, the effective scalar potential 
of the model is a function of the gluino bilinear only. 
Let us now suppose that we would like
to deal with the vacuum state of the model with a  nonzero value 
of the imaginary phase of the gluino condensate. For such a vacuum
we have 
\begin{equation} 
{\mbox{Im}}(\langle\lambda\lambda\rangle_k )=  N\Lambda^3\sin\left({2\pi
k\over N}\right)\not=0~. 
\nonumber
\end{equation} 
The imaginary
part of the gluino bilinear is a CP odd quantity. The non-zero vacuum value
of this  quantity signals a strong CP violation in the effective 
potential. This CP odd phase cannot be rotated away 
as there are {\em no} massless fermions in the
effective Lagrangian.  Thus, if the scalar
potential is written in terms of the gluino bilinear only, then it
cannot {\em simultaneously} describe the $N$ different vacua and yet preserve 
the strong CP invariance of the model. 
We must therefore generalize the effective superpotential 
(\ref{vy}) by including additional fields
such that the resulting effective superpotential respects the ${\bf Z}_N$ 
discrete symmetry, 
and is also consistent with the absence of the strong CP
violation. This can be done by introducing the  ``glueball'' 
order parameter fields in the 
effective Lagrangian. Thus, the observable effects of the CP odd phase in  
the gluino condensate should be  compensated by another  CP odd
quantity, the non-zero vacuum value of the field $ G_{\mu\nu}
{\widetilde G}^{\mu\nu} $. Moreover, this 
should also restore  the discrete ${\bf Z}_N$ symmetry. 

{}The ${\bf Z}_N$ symmetric effective superpotential 
has recently been proposed in \cite{gabad}. This superpotential involves two
chiral superfields $S$ and ${\cal X}$, where $S$ is the same as before,
whereas ${\cal X}$  is a new superfield whose lowest component is related 
(in a rather complicated way) to
the gluon bilinears $\langle GG\rangle$ and $\langle G{\widetilde G}\rangle$
\cite{far,gabad}. The corresponding scalar potential contains the gluino
bilinear via the lowest component of the  $S$ chiral superfield, and also
``glueballs'' via the lowest component of the ${\cal X}$ superfield.  Indeed,
the lowest component of ${\cal X}$ on mass-shell
is related to the
highest component of $S$, {\em i.e.}, 
to the gluonic operators discussed above. 
Likewise, the  F-component of ${\cal X}$ is related to 
the lowest component of
$S$, {\em i.e.}, to the gluino bilinear. The explicit
relations between these fields  are governed by an unknown K\"ahler potential
in the on-shell formulation of the model. 
However,  one can think of ${\cal X}$
as a superfield which introduces the necessary order parameters into the
description and  restores the ${\bf Z}_N$ symmetry of  
the superpotential (\ref {vy}). 
The effective superpotential for the 
$S$ and ${\cal X}$ superfields is given by 
\cite{gabad}: 
\begin{equation}\label{SX0}
 {\cal W}_{\small{eff}}=N S\left[\ln \left(N \Lambda^3\over S\right)+1
 -{\cal X}+{1\over N} \sum_n c_n e^{n N {\cal X}} \right]~.
\end{equation}
Here the coefficients $c_n$ are ${\cal O}(1)$ in the large $N$
limit, and satisfy the following
constraints\footnote{\label{foot}The coefficients $c_n$ are not uniquely 
determined within the effective Lagrangian approach, but should in principle
be fixed via the instanton calculus. If so, it is reasonable to expect that
the sum over the integer $n$ in (\ref{SX0}) should contain either only
non-negative or only non-positive values (albeit this cannot be seen solely on
the grounds of the effective Lagrangian arguments). Without loss of generality
we can then assume that $n\leq 0$ in (\ref{SX0}).}:   $\sum_n c_n=0$, and
$\sum_n n c_n =1$. The additional (compared with (\ref{vy})) terms in
(\ref{SX0}) containing the ${\cal X}$ superfield restore the   discrete ${\bf
Z}_N$ symmetry as well as CP invariance of the effective action
\cite{gabad}\footnote{The corresponding effective action is ${\bf Z}_N$
symmetric as long as the K{\"a}hler potential depends on ${\cal X}$ and ${\cal
X}^*$ via  ${\bf Z}_N$ symmetric combinations, such as ${\cal X}+{\cal X}^*$,
$e^{nN{\cal X}}$, $e^{nN{\cal X}^*}$, and so on.}. In particular, the ${\bf
Z}_N$ transformations read: $S\rightarrow S\exp(2\pi il/N)$, ${\cal
X}\rightarrow {\cal X}-2\pi il/N$, where $l$ is an integer.
Thus, the superpotential
(\ref {SX0}) describes the $N$ chirally asymmetric vacua  in SQCD\footnote{
Note that in the corresponding on-shell scalar potential ambiguities related to
the different branches of the logarithmic function are canceled as well.}.
These vacua are given by $(S,{\cal X})=(S_k,{\cal X}_k)$, where $S_k=N\Lambda^3
\exp(-{\cal X}_k)$, and ${\cal X}_k=2\pi i k/N$, $k=0,1,...,N-1$. 

{}Here we would like to point out that the effective superpotential 
(\ref {SX0}) can be regarded as the Veneziano-Yankielowicz superpotential (\ref{vy}) 
where the  
parameter $\Lambda^3$ has been promoted to the ${\cal X}$ dependent chiral
superfield $\Lambda^3({\cal X})$ \cite{gabad}:
\begin{eqnarray}
 \Lambda^3({\cal X}) \equiv \Lambda^3\exp \left(
 -{\cal X}+{1\over N} \sum_n c_n e^{n N {\cal X}} \right)~.      
\end{eqnarray}
Then the dynamics of the ${\cal X}$ field defines the phase of
$\Lambda^3({\cal X})$ (or, equivalently, 
the phase of the gaugino condensate) in 
accordance with (2),
and the superpotential (\ref {SX0}) can be rewritten in the following useful 
form:
\begin{equation}\label{SX00}  {\cal W}_{\small{eff}}=NS\left[
\ln \left(N\Lambda^3 ({\cal X})\over S\right)+1\right]~. \end{equation}  

{}The following notation will prove convenient in the subsequent discussions. Let
\begin{equation}\label{XX}
 X\equiv Ne^{-{\cal X}}~.  
\nonumber
\end{equation}
The effective superpotential $(\ref{SX0})$ can now be written as
\begin{equation}\label{SX}
 {\cal W}_{\small{eff}}=
NS\left[ \ln \left(\Lambda^3 f(X)\over S\right)+1\right]~.
\end{equation}
Here $f(X)$ is given by
\begin{equation}
 f(X)=X\exp\left[{1\over N}\sum_{n\ge 0} c_{-n} 
\left(X\over N\right)^{nN}\right]~,
\end{equation}
and has the following properties:
\begin{equation}
 f^\prime (X_k)=0~,~~~f(X_k)=X_k\equiv N\exp(2\pi ik/N)~.
\nonumber
\end{equation}
Note that $f(X)$ can be rewritten as a polynomial:
\begin{equation}
 f(X)=N \sum_{n\ge 0} {d_n\over {nN+1}} \left(X\over N\right)^{nN+1}~,
\end{equation}
where the sum is over non-negative integers, and the coefficients 
$d_n\sim 1$ are related to $c_n$'s and, as a result,  satisfy the
following constraints:
\begin{equation} \label{d}
 \sum_{n\ge 0} d_n=0~, ~~~\sum_{n\ge 0} {d_n\over {nN+1}}=1~.
\end{equation}
In particular, $d_0=\exp(c_0/N)=1+{\cal O}(1/N)$. 

{}It is not difficult to see that the superpotential (\ref{SX}) has ${\bf Z}_N$
discrete symmetry (with the transformations $S\rightarrow S\exp(2\pi
il/N)$, 
$X\rightarrow X\exp(2\pi il/N)$), and the corresponding vacua are given by
\begin{equation}
 S=S_k= X_k \Lambda^3~,~~~X=X_k~,~~~k=0,1,\dots,N-1~.
\end{equation} 
Thus, the superpotential (\ref{SX}) can be used to describe large $N$ BPS
domain walls  interpolating between two different chirally asymmetric 
vacua $(S,X)=(S_k,X_k)$ and $(S,X)=
(S_{k^\prime},X_{k^\prime})$ 
with $k^\prime=k+ 1$. Since the effective
superpotential  (\ref{SX}) is a truncation of the full effective Lagrangian
(which unlike (\ref{SX})  contains an infinite number of fields), it is
important to understand the applicability limits of this approach. Let the
center of a domain wall solution  (which depends only on the $z$ coordinate
and is independent of the $x,y$ coordinates)  be located at a point $z=z_0$.
Then, the effective Lagrangian approach should be applicable to study the
domain wall at  distances \begin{eqnarray}
|z-z_0|\gg d \propto {1\over N}~,
\end{eqnarray}   
where $d$ stands for the ``width'' of the wall which is vanishing in the large $N$ limit. 
This is the key simplification which allows us to explicitly construct 
BPS domain walls in the large $N$ SQCD.

\subsection{BPS Domain Walls in ${\cal N}=1$ Theories}

{}Before we turn to finding BPS domain walls in the large $N$ SQCD, 
in this subsection we would like to review some useful details concerning
generic properties  of domain walls in ${\cal N}=1$ supersymmetric theories.

{}Consider an ${\cal N}=1$ supersymmetric theory with 
one chiral superfield $X$. 
Let us assume that the corresponding K\"ahler potential does not contain
superderivatives. This is a reasonable assumption
for an effective theory in the large $N$ limit since
higher derivative terms
are expected to be suppressed by powers of $1/N$. 
Let ${\cal W}(X)$ be the superpotential 
in this theory such that the
F-flatness condition ${\cal W}_X=0$ has a discrete set
of non-degenerate solutions $X=X_a$, $a=1,\dots,N$.

{}In such a theory we expect presence of domain walls separating 
inequivalent vacua
$X=X_a$ and $X=X_b$, $a\not=b$.
Let $z$ be the spatial coordinate transverse 
to the wall. Then asymptotically
we have $X(z)\rightarrow X_a$ as $z\rightarrow-\infty$, and 
$X(z)\rightarrow X_b$ as $z\rightarrow+\infty$. Such domain 
walls may or may not be BPS saturated. 

{}The problem of finding BPS domain walls in four dimensional ${\cal N}=1$ 
supersymmetric theories is really a two dimensional problem as 
the coordinates $x,y$
along the wall play no role in the discussion. Thus, our problem 
effectively reduces to that
of finding BPS solitons in two dimensional massive ${\cal N}=2$ quantum 
field theories \cite {LG}. 
The corresponding BPS equation for the lowest component of the $X$ superfield
takes the following form: 
\begin{eqnarray}
\partial_z X^*(z)=e^{i\gamma} F^*_X,
\nonumber
\end{eqnarray}
where $F_X$ denotes the F-component of the $X$ superfield, and $\gamma$ 
is some constant phase (whose precise value will be given in a moment). 
If the K\"ahler potential 
contains no superderivatives, then $ F^*_X$ is related to the 
derivative of the superpotential via the following equation of motion:  $ g
F^*_X= W_X$, where $g\equiv g(X,X^*)$ is the K{\"a}hler metric defined
through the K{\"a}hler  potential ${\cal K}$ as $g={\cal
K}_{XX^*}$. Combining  the  BPS equation  and the equation of motion for 
$F^*_X$
one finds: $ g \partial_z X^* = \exp(i\gamma) {\cal W}_X $. Let us now turn to
the  calculation of the surface energy density of a domain wall
which satisfies  this relation. For a general supersymmetric theory with one
chiral superfield  and with no superderivatives in the K\"ahler potential
the energy of a  time-independent configuration is given by 
\begin{equation}
 E_{ab}={1\over 2} \int_{-\infty}^{+\infty} dz 
\left(g\vert\partial_z X\vert^2 +V\right)~,
\end{equation}  
where $V=g^{-1}\vert W_X\vert^2$ is the scalar potential of the theory. 
Note that $E_{ab}$ can be rewritten as follows: 
\begin{equation}
 E_{ab}={1\over 2} \int_{-\infty}^{+\infty} dz g^{-1} \left| g
\partial_z X^*-\exp(i\gamma)
 {\cal W}_X \right|^2 +{\mbox{Re}}\left(\exp(i\gamma)({\cal W}_{b}-
{\cal W}_a)\right)~,
\end{equation}
where $\exp(i\gamma)$ is an arbitrary constant phase. Note that $E_{ab}$ 
is independent of $\gamma$. For the choice $\gamma=\gamma_{ba}$ we obtain 
the bound $E_{ab}\geq
|{\cal W}_{b}-{\cal W}_a|$ which is the BPS bound. Here
\begin{equation}
\exp(-i\gamma_{ba})= {{\cal W}_{b}-{\cal W}_a\over |{\cal W}_{b}-{\cal W}_a|}~.
\end{equation}
The BPS solutions (for which $E_{ab}= |{\cal W}_{b}-{\cal W}_a|$) are those 
that
satisfy the following equation:
\begin{equation}\label{BPS}
 g \partial_z X^* = \exp(i\gamma_{ba}) {\cal W}_X
\end{equation}
subject to the boundary conditions $X(z)\rightarrow X_{a,b}$ as 
$z\rightarrow \mp\infty$.
Going back to the domain walls in four dimensions, we have the exact same 
BPS equation
(\ref{BPS}), and the tension of the domain wall is given by
\begin{equation}
 T_{ab}={1\over 8\pi^2}|{\cal W}_{a}-{\cal W}_b|~.
\end{equation}

{}Typically, one does not know the exact form of the K{\"a}hler metric $g$. 
However, for the case of a single superfield $X$ we are considering here 
one can still make certain statements about the corresponding soliton 
solutions. Thus, let $X$ be a function of a new coordinate $z^\prime$ 
such that it satisfies the following equation 
\begin{equation}\label{BPS1}
 \partial_{z^\prime} X^* = \exp(i\gamma_{ba}) {\cal W}_X
\end{equation}
with the boundary conditions $X(z^\prime)\rightarrow X_{a,b}$ as $z^\prime 
\rightarrow \mp\infty$. Suppose we are able to find the corresponding 
solution for $X(z^\prime)$. Next, consider the following change of variables:
\begin{equation}\label{diff}
 \partial_{z^\prime} z(z^\prime) = g(X(z^\prime),X^* (z^\prime))~,
\end{equation}
where $X(z^\prime)$ is the corresponding solution. Note that if we express 
$X$ as a function of $z$, then $X(z)$ will satisfy the original BPS equation 
(\ref{BPS}) with the corresponding boundary conditions. The change of 
variables (\ref{diff}) is simply a diffeomorphism which is one-to-one as 
long as the K{\"a}hler metric $g$ is non-singular. Throughout this paper 
we will 
assume that the corresponding K{\"a}hler metric $g$ is indeed non-singular. 
Moreover, we will always work in the coordinate system parametrized by 
$z^\prime$, but we will drop the prime for the sake of simplicity of the 
corresponding expressions. (This is effectively equivalent to the case where 
$g=1$.) Note that as long as the K{\"a}hler metric $g$ is non-singular,
the solution of (\ref{BPS1}) (with the appropriate boundary conditions) 
implies that the corresponding solution of (\ref{BPS}) exists. Moreover, 
this solution is BPS saturated.

{}In the next section we will see that the problem of
finding BPS domain walls in the large $N$ SQCD is reduced to 
that of 
finding BPS solitons in the so called $A_N$ 
Landau-Ginzburg model. The solitons in 
Landau-Ginsburg theories have been studied in detail 
\cite{LG,FMVW,CV}. Many of these models are integrable. The general
conditions for existence  of BPS solutions were first 
formulated in \cite{CV}, where exhaustive studies
of related issues were performed.
The superpotential in the $A_N$
Landau-Ginsburg model (in our notations) is given by 
\begin{equation}
{\cal W}\propto \Lambda ^3 \Big ( X-{X^{N+1}\over{N+1}}\Big )~.
\end{equation} 
There are $N$ distinct vacua 
in this model with $X=X_k=\exp(2\pi ik/N)$, $k=0,\dots,N-1$. The solitons
interpolating between the vacua $X_k$ and $X_{k^\prime}$ are BPS saturated
\cite{FMVW}.

\section{Large $N$ Domain Walls}

{}In this section we explicitly construct BPS domain walls in the 
large $N$ pure supersymmetric QCD
using  the model effective superpotential (\ref{SX}). We will be
looking for BPS solutions interpolating between neighboring vacua with phases
$k$ and $k^\prime=k+1$. 
Before we turn to the technical details of the 
corresponding equations,  we would like to make a few observations which will
lead to  important simplifications, and will elucidate some features of the
system we are dealing with.

{}To begin with, let us note that the following
relation holds for the vacuum state labeled by the phase $k$: 
$S|_k=\Lambda^3
f(X)|_k$. This relation is nothing but the definition of the
gluino condensate with the corresponding phase set by the vacuum value
of the $X$ superfield. It is useful to introduce the chiral superfield
$NS/\Lambda^3 f(X)$. In all the vacua this superfield takes the same value
equal $N$. 
Let us now concentrate on interpolation 
between a pair of nearest-neighboring vacua. In this case
the relative change in the gluino bilinear 
is of order $1/N$. 
Hence, the chiral
superfield $N S/
\Lambda^3 f(X)$ can only deviate from its vacuum value (which equals $N$) by
a quantity of order 1. Thus, one can introduce the following parametrization: 
\begin{eqnarray}
{NS\over \Lambda^3 f(X)}=N\left(1- {\Sigma\over N}\right)~,
\label{sig}
\end{eqnarray}
with $\Sigma $ being a new chiral superfield.
Thus, the $S$ superfield is defined via the $X$ and $\Sigma$ fields
as far as the nearest-neighboring vacuum interpolation is concerned.
Note that both real and imaginary parts of $\Sigma$ are at most of 
order 1 in the large $N$ limit. Substituting the expression
(\ref {sig}) into the superpotential
(\ref {SX}), one finds
\begin{eqnarray}
{\cal W}_{\small{eff}}= N \Lambda^3 f(X) \left[ 1- {\Sigma^2\over N^2 }
+{\cal O} \left ( {1\over N^3} \right )  \right]~. \nonumber
\end{eqnarray}
This is the superpotential describing the variation of the order parameter 
during the transition between nearest-neighboring vacua.
The superfield $\Sigma$ enters this superpotential in a
subleading order in $1/N$.   
Thus, it is the superfield $X$ which is left in the leading order, 
and which should describe 
domain walls between the adjacent vacua in the large $N$
limit. Neglecting higher order terms, the superpotential
is therefore given by: 
\begin{eqnarray}
{\cal W}_{\small{eff}}\simeq N \Lambda^3 f(X)~. 
\label{superX}
\end{eqnarray}
Before we go any further, notice that this description is valid only for 
a pair of adjacent vacua, and {\em a priori} 
the approximations made in (\ref {superX})
cannot be justified for a pair of vacua for which $|k-k'|\sim N$. 

{}Let us now turn to the K\"ahler potential  describing
transitions between a pair of nearest-neighboring vacua.
Generically, the K\"ahler potential is a function of the entire tower of SQCD
composite fields and their superderivatives. It is reasonable to expect,
however, that in the large $N$ limit the superderivative terms are
suppressed by extra powers of $1/N$.
Thus, we assume the following form for the K\"ahler potential $K=K(\Psi_i,
\Psi_i^*, X, X^*)$. Here, $\Psi_i$  denote  all the fields  which are
generically present in the theory, 
but which do not appear in the large $N$ 
effective superpotential (\ref {superX}) describing the
nearest-neighboring
transitions. Notice that this set of fields includes the superfield
$\Sigma$ as well (in accordance with (\ref {sig}) one of $\Psi_i$ can 
be defined as $\Psi_l\equiv N-\Sigma$).
Here we are normalizing all the $\Psi_i$ fields so that in the large $N$ limit
they  scale as $\sim N$ (just as is the case for $S$ and
$X$). This, in particular, implies that the corresponding components of the 
K{\"a}hler metric are (at most) of order 1 in the large $N$ limit. 
Using the K\"ahler potential and the 
superpotential (\ref {superX}) one can write
down both the equations of motion and the BPS equations. 
Let us start with  the equations of motion first. For 
simplicity of presentation we write down only some part of the system
of equations of motion relevant for our purposes here. 
We must make sure, however, that eventually
all the equations of motion are satisfied. 
Since
there are no superderivatives in $K$, one can write 
the following equations of
motion in the leading order in $1/N$:
\begin{eqnarray}
\partial (g_{\Psi_i \Psi_j^*} \partial \Psi^*_j)+ \partial
(g_{\Psi_i X^*} \partial X^*)\simeq 0~, \nonumber \\
g_{\Psi_i\Psi_j^*} F_{\Psi_j}^*+ g_{\Psi_i X^*} F_X^*\simeq 0~,
\nonumber \\
g_{X X^*}F_X^* + g_{X \Psi_i^*} F_{\Psi_i}^*\simeq
{\cal W}_{X}~,
\label{EOM}
\end{eqnarray}
where the sign "$\simeq$" denotes equality 
in the leading order of the large $N$ expansion. 
We are going to argue below that the fields $\Psi_i$
are irrelevant for the description of the domain walls
interpolating between adjacent vacua. In other words, 
we will argue that the domain walls between the neighboring vacua are 
described by the superfield $X$ alone. To see how this 
comes about let us recall that 
the discrete ${\bf Z}_N$ symmetry is spontaneously broken in the
theory. Thus, there are domain walls separating adjacent vacua. 
The question 
we are dealing with is whether these walls are BPS saturated or not. In
other words, there should always exist a wall configuration which would 
satisfy
the equations of motion. However, the very same configuration might or might
not  satisfy the corresponding BPS equations. Let us start with the equations
of motion defined in (\ref {EOM}). 
Given the boundary conditions  $\partial \Psi(\pm \infty)=0$ and
$\partial X(\pm \infty)=0$,
and also taking into account non-singularity of the K\"ahler metric, 
one can rewrite the first equation in (\ref {EOM}) as follows:
\begin{eqnarray}
g_{\Psi_i\Psi^*_j}\partial\Psi^*_j+g_{\Psi_i X^*}\partial X^*\simeq 0~.
\label{first}
\end{eqnarray}
While making a transition from one vacuum state to the 
neighboring one along the solution, the relative change in the 
fields $\Psi_i$ and $X$ should be at most of order $1/N$. 
We  therefore will be looking for solutions of the form 
$\Psi_i=\Psi_{i0}+\Psi'_i$, where $\Psi_{i0}$ (which, if non-vanishing, 
scale as $\sim N$) denote some constants
and $\Psi'_i$ set the coordinate dependence of the 
fields in the nearest-neighbor transitions. For instance, as we have
already mentioned, the $\Sigma$ field enters 
the equations via  $\Psi_l=N
(1-\Sigma/N)$. Likewise, the large $N$ behavior of the $X$ superfield  
is set by the expression $X=N(1-\zeta/N)$, where both the real
and imaginary parts of $\zeta$ are at most  
of order 1.  Given these relations  the coefficients
of the K\"ahler metric, which also should be varying by
an amount proportional to $1/N$, can be written in the
following form:
\begin{eqnarray}
g_{\Psi_i \Psi^*_j}|_{\rm solution}=C_{ij}
+{\cal O} \left ({\Psi'\over N},~{\zeta \over N}   \right )~,
\nonumber \\
g_{\Psi_i X^*}|_{\rm solution}=C_i+{\cal O}\left (
{\Psi'\over N},~{\zeta\over N} \right)~,   \label{metric}
\end{eqnarray}
where $C_{ij}$ and $C_i$ are some constants of (at most) order 1.
Substituting these relations into (\ref {first}) and taking into account
the boundary conditions for the $X$ superfield 
as well as those for the $\Psi'_i$ fields,  $\Psi'_i (\pm
\infty)=0$, one finds that (\ref {first}) admits  no 
solutions but the trivial one: $\Psi'_i=0$ provided that 
$g_{\Psi_i X^*}|_{\rm solution}\propto {\cal O} ({\zeta/N})$, that is,
$C_i\simeq 0$.
Note that since domain walls must exists (regardless of whether they are
BPS saturated or not), the last condition on the K{\"a}hler potential {\em
must} be satisfied. On the other hand, as we will show in a moment, once
this condition is satisfied, the corresponding walls {\em do} saturate the
BPS bound. Next, to solve for $\zeta$, or equivalently $X$, 
one should use the rest of the equations in (\ref {EOM}).
Let us turn to the second equation in (\ref {EOM}). 
As we have shown 
above the off-diagonal components  of the metric $g_{\Psi_i X^*}$
should be of the subleading order on the solution. Thus, in the leading
order of the large $N$ expansion the 
second equation in (\ref {EOM}) turns into the relation
$g_{\Psi_i \Psi_j^*}F^*_{\Psi_j}|_{\rm solution}\simeq 0$. Consider the
trivial solution $F_{\Psi_i}\equiv 0$. This is precisely the solution we
would like to consider for the purpose of finding BPS domain walls. Indeed,
this solution implies that the set of BPS equations for $\Psi_i$, namely,
$\partial \Psi_i\propto F_{\Psi_i}$ is automatically satisfied (recall that   
$\partial \Psi_i\equiv 0$ as a result of the first equation in (\ref{EOM})).

Finally, let us discuss the third equation in (\ref {EOM}).
Given the results of the first equation  one finds in the leading
order: 
\begin{eqnarray} 
g_{X X^*}F_X^* \simeq  {\cal W}_{X}~.
\label{F_X}
\end{eqnarray}
Let us now turn to the corresponding BPS equation. 
In accordance with our discussions
presented above  
the only nontrivial BPS equation is that
for  $X$,  $\partial_z X^* (z) =\exp(i\gamma_{k+1,k}) F^*_{X}$, where 
the phase $\gamma_{k+1,k}$ was defined in the previous section. 
As the next step,
one can use the
equation of motion (\ref {F_X})  to derive:  
\begin{eqnarray} g_{X
X^*}\partial_z X^*\simeq \exp(i\gamma_{k+1,k}){\cal W}_{X}~. 
\end{eqnarray}
Below, we will assume that the corresponding K{\"a}hler metric
$g_{X X^*}$ is non-singular. Then, as we reviewed in the previous section,
we can absorb the K{\"a}hler metric into a redefinition of $z$. (More 
precisely, we will absorb $g_{X X^*}/\Lambda^2$ 
which is dimensionless.) In fact, in the large $N$ limit
this procedure is equivalent to the absorption of a finite
constant in the redefinition of the variable $z$, or, equivalently,
to the redefinition of the width of the domain wall by a finite number.
The  corresponding BPS equation reads:
\begin{equation}\label{bps}
 \partial_z { X}^* \simeq \exp(i\gamma_{k+1,k}) \Gamma 
 { f}^\prime ({ X})~, 
\end{equation}
where we have introduced the inverse width 
\begin{equation}
 \Gamma=d^{-1}= N\Lambda~.
\end{equation}
The boundary conditions for ${ X}(z)$ are given by: 
${X}(-\infty)=X_k$, ${X}(+\infty)=X_{k+1}$.

Thus, the system of BPS equations and equations of motion for  
this particular case reduce formally to the system 
described by the superpotential
\begin{eqnarray}
{\cal W}\simeq N \Lambda^3 f(X)
\label{W}
\end{eqnarray}
amended by the $S$ field defining condition 
$$
S= \Lambda^3 f(X)~(1+{\cal O}({1/N^2}))~.
$$
In addition, the values of the all other fields $\Psi'_i$ 
(including $\Sigma$) on the solution vanish in the leading order
of the $1/N$ expansion. 
These relations are supposed to be satisfied on the particular BPS solution
we are dealing with. In the following subsections
we will discuss the solution to the BPS equation  (\ref {bps})  
and present the explicit form of the corresponding BPS domain wall.

\subsection{A Simple Example}

{}Before solving the BPS equation (\ref{bps}) for the most 
general form of $f(X)$, 
in this subsection, for illustrative purposes, we will consider a simple
example where we take the values of the numerical coefficients $d_n$ as
follows: $d_0=-d_1=1+1/N$, and all the 
other $d_n$ ($n>1$) coefficients are
${\cal O}(1/N)$. Then in the leading order in $1/N$ one finds: 
\begin{equation}\label{ff}  
f(X)\simeq d_0 \Big ( X-{N\over N+1}
\left(X\over N\right)^{N+1}\Big )~. 
\end{equation} 
Note that the effective
superpotential (\ref{W}) in this case reduces to that of the $A_N$ model
discussed in the previous section.

{}Next, let us solve the BPS equation (\ref{bps}) in the large $N$ limit 
for $f(X)$ given by (\ref{ff}). 
(In this subsection we will closely follow the corresponding discussion in \cite{gia}.)
Let ${X}=X_k Y$. Then we can rewrite (\ref{bps}) in terms of $Y$:
\begin{equation}\label{Y} 
 N\partial_z Y^*\simeq-i\exp(-\pi i/N)\Gamma\left[1-Y^N\right]~.
\end{equation}   
The boundary conditions on $Y$ read: 
\begin{equation} 
 Y(z\rightarrow-\infty)=1~,~~~Y(z\rightarrow+\infty)=\exp(2\pi i/N)~.
\end{equation}
Thus, in the large $N$ limit ${\mbox{arg}}(Y)$ changes by a small amount, 
namely, from $0$
to $2\pi/N$. It is convenient to parametrize $Y$ as follows:
\begin{equation} 
 Y=(1-\rho/N)\exp(i(\phi+\pi)/N)~.
\end{equation}
Here $\rho$ and $\phi$ are real. The boundary conditions then read:
\begin{equation} 
 \rho(z\rightarrow\pm\infty)=0~,~~~\phi(z\rightarrow\pm\infty)=\pm\pi~.
\end{equation}
Then (\ref{Y}) becomes a system of the following first order differential 
equations: 
\begin{eqnarray}
 &&\partial_z \phi=\Gamma \left[1+\exp(-\rho)\cos(\phi)\right]~,\\
 &&\partial_z \rho=-\Gamma \exp(-\rho)\sin(\phi)~.
\end{eqnarray}
Here we are taking the large $N$ limit and only keeping the leading terms. 
In particular, we have taken into account that $(1-\rho/N)^N\rightarrow
\exp(-\rho)$ as $N\rightarrow\infty$.

{}The above system of differential equations can be
integrated. Note that these equations do not explicitly contain $z$. 
This implies that
if $\phi(z)$ and $\rho(z)$ give a solution with the appropriate boundary 
conditions,
so will $\phi(z-z_0)$ and $\rho(z-z_0)$ for any constant $z_0$. 
(This is simply the statement that the system possesses translational 
invariance in the $z$ direction.) The general solution is given by
\begin{eqnarray} \label{solutions}
 &&\cos(\phi)=(\rho-1)\exp(\rho)~,\\
 &&\int_{\rho(z_0)}^{\rho(z)} d\xi \left[\exp(-2\xi)-(1-\xi)^2
 \right]^{-{1\over 2}}=-\Gamma |z-z_0|~,
\end{eqnarray}
where $\rho(z_0)=\rho_0 (\approx 1.278)$ is the solution of the following 
equation:
\begin{equation} 
 (\rho_0-1)\exp(\rho_0)=1~.
\end{equation}
Note that $z_0$ is the center of the domain wall. These solutions
have characteristic for a domain wall behavior. The amplitude $\rho$
changes from zero at $-\infty$ to zero at $+\infty$ with a bell-shaped
extremum  at $z=z_0$. Likewise, the phase $\phi$ changes monotonically
from $-\pi$ at $-\infty$ to $+\pi$ at $+\infty$ going through the origin 
at $z=z_0$. The width of this wall $d\sim 1/N\Lambda$ vanishes in the large
$N$ limit.

\subsection{The General Case}

{}In this subsection we construct the BPS domain walls for the most 
general form of $f(X)$. In fact, as we will see in a moment, the
corresponding BPS equations are actually reduced to  those discussed in the
previous subsection. That is, for the most general effective superpotential
${\cal W}$ in (\ref{W}) the problem of the large $N$ domain walls can be
reduced to that of   finding BPS solitons in the large $N$ $A_N$
Landau-Ginsburg theory. 

{}To begin with, let us simplify the form of the effective superpotential as follows. 
Let us make the following change of variables:
\begin{equation}
 f(X)={\widetilde f}({\widetilde X})\equiv {\widetilde X} - 
 {N\over N+1} \left({\widetilde X}\over N\right)^{N+1}.
\end{equation}
This transformation preserves the ${\bf Z}_N$ symmetry:
in terms of the new variable ${\widetilde X}$ we still have $N$ 
non-degenerate vacua ${\widetilde X}=X_k=N\exp(2\pi ik/N)$, $k=0,1,\dots,
N-1$. Also, the above non-linear change of variables is non-singular
as long as the coefficients $d_n$ satisfy constraints (\ref {d}).
This implies that the resulting K{\"a}hler metric 
${\widetilde g}({\widetilde X},
{\widetilde X}^*)$ is non-singular as well 
(provided that the original K{\"a}hler 
metric $g(X,X^*)$ was  non-singular). Therefore, as we reviewed in the previous
section, we can absorb the K{\"a}hler metric into a redefinition of $z$. (More 
precisely, we will absorb ${\widetilde g}/\Lambda^2$ 
which is dimensionless.) The 
corresponding BPS equation reads:
\begin{equation}\label{bps1}
 \partial_z {\widetilde X}^* =\exp(i\gamma_{k+1,k}) \Gamma 
{\widetilde f}^\prime ({\widetilde X})~, 
\end{equation}
where the phase $\gamma_{k+1,k}$ is the same as in the previous subsection.
The boundary conditions for ${\widetilde X}(z)$ are given by: 
${\widetilde X}(-\infty)=X_k$,
${\widetilde X}(+\infty)=X_{k+1}$. Note that the BPS equation (\ref{bps1})
is the same as that derived in \cite{gia} using a more indirect construction.
In particular, it coincides with the BPS equation we just solved in the previous subsection 
for a simple example corresponding to the $A_N$ superpotential. 
Thus,
the problem of BPS domain walls in the large $N$ SQCD is indeed reduced
to that of finding BPS solitons in the large $N$ $A_N$ Landau-Ginsburg theory.
In fact, this is a generic property of ${\bf Z}_N$ symmetric Landau-Ginsburg theories -
they are all related via non-linear change of variables which is non-singular. 
This implies that if the K{\"a}hler metric is non-singular for one choice of
variables, it is non-singular in the transformed variables as well. Since the
induced K{\"a}hler metric can always be absorbed into the redefinition of $z$,
this allows us to solve the BPS equations exactly (in the large $N$ limit) in
the suitable coordinate ``frame''. 

{}For completeness we note that the shape of the gaugino condensate is given by
\begin{equation}
 S=\Lambda^3 f(X)=\Lambda^3 {\widetilde f} ({\widetilde X})\simeq \Lambda^3
{\widetilde X}~. \end{equation}
Thus, in the large $N$ limit the gaugino condensate is controlled by the VEV of the 
chiral superfield ${\widetilde X}$. This important simplification is the key
observation of \cite{gia}. In particular, as we explain in section IV, the
additional singlet superfield used in \cite{gia} can be related to
${\widetilde X}$. This point is central to the fact that the indirect
construction of \cite{gia} gives BPS domain walls in the large $N$ SQCD, and
not in some other theory.

{}The tension of the above domain walls is given by:
\begin{equation}
 T_D={N\over 4\pi} \Lambda^3~.
\end{equation}
The width of these domain walls is $\sim \Gamma^{-1}$ which goes as $\sim 1/N$.
Thus, these domain walls are infinitely thin in the large $N$ limit.
Moreover, they are BPS saturated. 
Note that there are
other domain walls in large $N$ SQCD \cite{gia}, namely,  
those with $k^\prime \not=k\pm 1$. We refer the reader to 
\cite{gia} for details.

\subsection{Domain Walls and the Three-form Supermultiplet}\label{pbrane}

{}The large $N$ SQCD domain walls found in the previous sections can be viewed
as  membranes which break  half of SUSY generators and are embedded in a 
$(3+1)$-dimensional SQCD background. If so, a
three-form supermultiplet  \cite{townsend} must couple to such a membrane to
have  world-volume $\kappa$-symmetry \cite{duff}\footnote{ 
As usual, $\kappa$-symmetry eliminates half of the world-volume fermionic
degrees of freedom, which is necessary for having a supersymmetric 
formulation of the membrane action.}. 
If the domain walls are indeed 2-branes (membranes), we 
must be able to identify the corresponding three-form supermultiplet with 
the appropriate $(3+1)$-dimensional ``bulk'' 
physical fields of  SQCD. 
In fact, in the following we will argue that after gauge fixing the  
three-form superfield reduces to component fields, which in a given chirally 
asymmetric vacuum describe the corresponding gluon and/or gluino composites
of SQCD. 

{}Before we turn to this identification, let us make the following remark. 
Consider the domain wall separating two neighboring 
chirally asymmetric  vacua. 
In each of these vacua there are  SQCD fields with a well defined particle 
interpretation in terms of colorless composite bound states of gluons and/or gluinos.
We would like to show that the SQCD fields in each of 
the two vacua consistently 
couple to the membrane world-volume. 
Note that the corresponding couplings for these two vacua are different. 
A consistent ``sewing'' of these couplings is guaranteed by the fact 
that the domain wall is a solution (interpolating between these two vacua) 
of the order parameter effective action (\ref{SX0}). 
In the following we will therefore identify the components of the 
three-form supermultiplet with 
SQCD colorless fields in a given chirally asymmetric vacuum.  

{}Let us start with the supersymmetric Green-Schwarz action 
for a 2-brane (membrane) embedded in the $(3+1)$-dimensional target space. 
The membrane action ${\cal S}_m$ includes the Wess-Zumino term corresponding to the
three-form coupling to the world-volume of the membrane:
\begin{eqnarray}
 {\cal S}_m=T_D \int d^3\sigma \left ( -{1\over 2}  \sqrt{-{\cal G}} 
{\cal G}^{ab}E_a^i E_b^j \eta_{ij} +{1\over 2} \sqrt{-{\cal G}} 
 +{1\over 3!} \epsilon^{abc}E_a^A E_b^B E_c^C
 B_{ABC}\right),
\label{supermembrane}
\end{eqnarray}
where $\sigma_a$ ($a=0,1,2$) denote the membrane world-volume coordinates, 
and ${\cal G}_{ab}$ is the induced metric on the
world-volume. Here the pull-back $E_a^A = \partial_a
Z^M(\sigma) E_M^A$, $Z^M\equiv (x^{\mu}, \theta^{\alpha})$ are 
the target superspace coordinates, and $E_M^A$ denotes the 
supervielbein, where $A=(i,\alpha)$ are the tangent superspace indices.   
The constraints \cite {townsend,ovrut} imposed by $\kappa$-symmetry on 
the three-form superfield $B_{ABC}$ can
be solved, and the remaining independent components of $B_{ABC}$ can 
be combined into the following real tensor 
superfield:
\begin{eqnarray}
U=&&B+i\theta \chi -i {\bar \theta} {\bar \chi}+{1\over 16}\theta^2 {A^*}+
{1\over 16} {\bar \theta}^2 A+{1\over 48 }\theta \sigma^\mu {\bar
\theta} \epsilon_{\mu\nu\lambda\rho}C^{\nu\lambda\rho}+ 
\nonumber \\
&&{1\over 2} \theta^2 {\bar \theta} \left ( {\sqrt{2} \over 8}{\bar
\Psi} +{\bar \sigma}^\mu \partial_\mu \chi \right )+
{1\over 2}{\bar  \theta}^2 \theta  \left ( {\sqrt{2} \over 8}
\Psi - \sigma ^\mu \partial_\mu {\bar \chi }\right )+{1\over 4}
\theta^2 {\bar \theta^2} \left ( {1\over 4} \Sigma -\partial^2 B\right )~.
\label{U}
\end{eqnarray} 
We would like to point out  
that the supermembrane action by itself 
gives rise to the real tensor superfield 
(\ref {U}) only defined up to a shift by
a linear supermultiplet $L$ satisfying ${\bar D}^2L=D^2L=0$ \cite 
{townsend,ovrut}. 
If the membrane is  considered in empty space, then this shift, 
$U\rightarrow U-L$, can 
always be used to remove  half of the components in (\ref {U}). 
However, in our case the membrane is
embedded in a space where composite colorless SQCD fields live. Some of these
fields, as we are going to see shortly, couple to the membrane.    
Thus, the total classical action of the system 
is the sum of the membrane action
${\cal S}_m$ and the effective action for the bulk fields 
\begin{equation}
{\cal S}={\cal S}_m+{\cal S}_{bulk}~.
\end{equation}
The action for the bulk fields ${\cal S}_{bulk}$ is a rather 
complicated object and
generically consists of an infinite number of fields. However, it was shown 
in \cite {far} that lowest-spin SQCD excitations (spin-zero glueballs, 
gluino-gluino mesons and their fermionic superpartners)  
can be  described by the components of   
the three-form supermultiplet $U$. It was also argued in \cite
{far} that  the  shift $U\rightarrow U-L$  is {\em not}
a symmetry of the bulk effective Lagrangian ${\cal S}_{bulk}$. Thus, all the 
components in (\ref {U}) should be retained as physical ones. 
If so, one can identify the fields in (\ref{U}) 
with the bulk composite massive fields of SQCD \cite {far}.   
Thus, the three-form field $C_{\mu\nu\lambda}$
is just the ``magnetic'' dual of the Chern-Simons current of 
SQCD, and is related
to the operator $G\widetilde G$ via
$\epsilon_{\mu\nu\lambda\rho}\partial^{\mu}C^{\nu\lambda\rho}\propto
G{\widetilde G}$. (Note that $C_{\mu\nu\lambda}$ is a {\em massive} field in 
this approach with one physical pseudoscalar degree of freedom \cite {far}.)
The two Weyl fermion states $\chi $ and $\Psi$ in (\ref {U})
correspond to $(l,s)=(0,1/2)$ and $(l,s)=(1,1/2)$
gluino-gluon bound states, respectively (where $l$ and $s$ denote the orbital 
momentum and spin respectively). The $A$ field
denotes the $\lambda\lambda$ state, whereas
the $B$ field is related to the $\Sigma$ field (on-shell) which 
corresponds to the scalar glueball given by the $G_{\mu\nu}^2$ composite 
operator.  

To summarize, there are four real massive scalar degrees of freedom  
corresponding to the scalar and pseudoscalar gluino-gluino composites, and
the scalar and pseudoscalar glueballs. In addition,  
there are four real (two Weyl)
fermionic degrees of freedom corresponding to the gluino-gluon composites
with different orbital momenta. These states are present in the
superfield (\ref {U}) which necessarily  appears in the supersymmetric
formulation of the membrane action. 

One might wonder now what is the relation between the bulk effective action 
${\cal S}_{bulk}$ and the order parameter effective action 
for two chiral superfields $S$ and ${\cal X}$ 
used previously. As we mentioned above, ${\cal S}_{bulk}$ depends in general
on an infinite number of superfields corresponding to different spin-orbital
excitations of SQCD. If one restricts consideration to the lowest-spin states
only, then those states precisely fit in the supermultiplet (\ref {U}) 
\cite {far},
so that ${\cal S}_{bulk}={\cal S}_{bulk}(U,...)$.
On the other hand, it has been  known for some time 
\cite{gates,derendinger} that the field content of the 
three-form supermultiplet $U$ can be
rearranged into two chiral superfields. This property was used in 
\cite{derendinger,far}
to rewrite the three-form supermultiplet $U$ in terms of  two chiral
superfields  which in our case, in a given chirally asymmetric vacuum, 
correspond to 
the chiral superfield $S$ and some function of 
the chiral superfield ${\cal X}$ \cite{gabad}. 

Before we turn to the next section we would like to comment on 
SQCD strings and some related phenomena.
There are a number of indications, as we discussed in 
the introduction, that in the large $N$ limit
QCD (or SQCD) can be viewed as some  non-critical string theory.  
Thus, one might  be able to describe 
all the SQCD states in the large $N$ limit 
as  string excitations, without actually referring
to the fundamental theory (SQCD). The question we  address  
here is whether some of the SQCD states discussed above
could indeed be identified
with excitations of strings present in the ``bulk''. 
In the  string picture glueballs are identified with excitations of 
a closed non-critical string: the scalar glueball 
in the SQCD string theory 
would correspond to the string ``dilaton'' \cite {ShifmanMigdal}.
The pseudoscalar glueball is the analog of the ``axion'' and, 
by supersymmetry, a gluino-gluon bound state corresponds to the 
``dilatino''. These states, as we discussed above,
along with the gluino-gluino bound states,
are combined in one massive tensor supermultiplet,
similarly to what can  happen in superstring theories at
the non-perturbative level (see discussions in the second reference in 
\cite {derendinger}).  
Note that the closed string itself, whose excitations 
we identify with the  glueballs,  can be viewed as a limit
of a closed membrane with the topology of a torus.  
This is consistent with the results of \cite{g}
where a part of the low-energy glueball spectrum of QCD 
was calculated in a closed bosonic membrane model 
and a good agreement with the lattice QCD predictions 
was found\footnote{In  
\cite {g}, because of computational difficulties, the spectrum of glueballs
was actually 
calculated for a bosonic membrane with the topology of a sphere.}. 
These observations might  provide some hints toward 
understanding the string theory of strong interactions.

Finally, we would like to comment on the  
possibility of  SQCD open strings ending on the domain walls. 
If the domain walls we found are to be 
interpreted as D-branes, there should be open strings ending on the 
brane \cite {witten}.
Thus, in the D-brane picture, 
one should in principle be able to identify the $U(1)$ gauge field in 
the membrane world-volume action \cite {Pol}. One might wonder how this 
$U(1)$ gauge field arises in the field theory context if the 
open strings we discuss are viewed  as
SQCD chromoelectric  flux tubes. An example of a wall on which 
the flux tube can end was  constructed in \cite{wall}. 
In that case, the corresponding theory is in the confining phase outside 
of the wall. Thus, all chromoelectric  fields  
are squeezed into flux tubes outside of the wall.
It is energetically favorable for the tubes to 
end on the wall and to spread the flux  into  the wall interior
if the gauge theory inside the wall is in the  Higgs phase \cite {wall}. 
Thus, the original gauge group
of the confining theory, being  Higgsed inside the defect, could support
flux tubes ending on the wall. 
The Higgs phase inside the defect can support 
the  residual unbroken $U(1)$ gauge group in the wall. 
In the large $N$ limit the wall becomes  infinitely thin. 
Thus, the system of the flux tube and the wall can be  regarded in that limit
as an SQCD string ending on a D-brane. The 
world-volume  $U(1)$ field in that case can be interpreted as 
a collective solitonic excitation in the soliton picture for the wall, 
or, alternatively,  as a mode of an  open string ending on the D-brane, 
if the D-brane picture is adopted.  Recently, some  
features of the system of $N$ such domain walls with flux tubes 
were discussed in \cite {ahs}. It would be interesting to study these
issues further.

\section{Relations to Previous Studies} \label{relation}

Surprisingly enough, the solution of BPS equations as well as superpotential
(\ref {ff}) coincide with the ones derived in \cite{gia} within a
completely different approach. As we will argue below, this coincidence is
not an accident. In \cite{gia} BPS domain wall solutions were found
in the theory obtained by integrating out heavy states in 
SQCD with $N_f$ heavy quark flavors. 
The method of gaining an analytic
control over BPS equations in pure SQCD by adding flavors and then
integrating them out was used earlier in a number of papers, and, in that
respect, was not new in  \cite {gia}.
However, there is a crucial point 
which makes conclusions of  \cite{gia} so different from other approaches. 
Let us explore this difference in somewhat more
detail. As is well known, the supersymmetric QCD with $N_f<N$ flavors
possesses an anomaly free R-symmetry under which flavors carry the charge
${N_f- N\over N_f}$. Adding a constant mass term for ``mesons'' $M$  in the
supersymmetric Lagrangian
\begin{equation}
 m{\rm Tr}(M)
\end{equation}
explicitly breaks this symmetry. However, in the massive theory there still is 
an anomaly free discrete  ${\bf Z}_{N}$ symmetry.   This group acts on the
gaugino bilinear, and, in fact, coincides in that respect with the anomaly
free ${\bf Z}_N$ symmetry of pure SQCD. 
Moreover, spontaneous breaking of this
discrete symmetry by meson VEVs should produce domain walls which after taking
the limit $m \rightarrow \infty$ are expected to transform 
continuously into the domain
walls of pure SQCD.  This was the method used in \cite{smilga}.
It was shown in \cite{smilga} that within this approach 
the BPS walls do exist for sufficiently small values of $m$, however, above
some critical $m_*$ the walls cease to exist. Thus, one would seemingly not
have BPS domain walls in pure SQCD which is  expected to be recovered in the
limit $m\rightarrow\infty$.   Although {\em per s{\'e}} there is nothing 
wrong with
solving equations for fixed $m$ 
and then extrapolating them into the large $m$
region, it seems to us that the domain wall solutions obtained this way
cannot be related to the walls in pure SQCD discussed in \cite{wall} and
explicitly constructed in the present work. The reason for this is as follows. 
The heavy mesons which are being integrated out in this approach 
transform under the discrete symmetry 
which also acts on the degrees of freedom 
of the corresponding low-energy theory.
If so, a straightforward decoupling of these fields  cannot be justified  for
the problem of finding BPS domain walls of the corresponding low-energy theory.
Indeed, it was  shown by Kogan, Kovner and Shifman 
in \cite {flux} that  when the heavy fields
are charged under the discrete symmetry (which is subsequently broken
spontaneously) the classical solutions that interpolate between different
vacua cannot be adequately described by simply integrating 
heavy fields out and solving equations of the
corresponding  effective low-energy theory. This inadequacy arises  due to
``cusp singularities''  which emerge in the low-energy theory once the
fields charged
under the discrete symmetry are integrated out \cite{flux}. 
Now, in SQCD with flavors and the regular mass term,   mesons necessarily
transform under the remaining ${\bf Z}_{N}$ discrete symmetry group, 
and, thus, the
``cusp crossing'' will always occur once the mesons are integrated out. 
In this case, albeit one can
consistently solve the BPS equations and find  the wall  
solutions for a finite mass $m$ \cite {smilga}, 
this method  would not  allow
one to account for the BPS walls of  pure supersymmetric
QCD which is reached in the 
$m\rightarrow \infty$ limit.
In order to avoid this difficulty, the  mesons which are being integrated
out should not transform under the discrete  symmetry. This is possible 
to accomplish if the
mass term is not a constant, but rather is promoted into a {\it dynamical
field}. This was the key observation  of \cite{gia}. Consider SQCD
with $N_f=N$ flavors. This theory has an anomaly free $U(1)_R$-symmetry under
which mesons are neutral. 
To avoid the presence of ``cusp singularities'', one can 
postulate that
the mass term for ``mesons'' 
arises due to  the VEV of a new chiral superfield $X$, and takes the form
$\Lambda  X{\rm Tr}(M)$. In
this case there is a discrete ${\bf Z}_{N}$ 
symmetry under which the  $X$ field and gauginos 
transform but mesons are neutral. 
This group, in fact, is  a subgroup of the 
anomaly free $U(1)_R$. Now, if the mesons acquire large masses due to the VEV
of the $X$ superfield, then these mesons can be safely integrated out
without encountering cusp singularities discussed above. 
Thus, one should be able to describe SQCD domain walls in this case. 
The necessary
condition is that one should work in the region where $X$ is large. For this,
one can add a small ${\bf Z}_N$-conserving term in the superpotential. The simplest
one is $X^{N+1}/(N + 1)$.  In this  case, after integrating mesons out, one
recovers (\ref {ff}). Another indication of a  
deeper connection between $X$ in
(\ref {SX}) and the one in \cite {gia} is the fact that in the
above context $X$ essentially acts as a dilaton (plus axion) field. This can
be seen in a number of ways. The simplest one is to point out that the VEV of
$X$ spontaneously breaks an anomalous $U(1)$-symmetry (the Peccei-Quinn
symmetry \cite {PQ}) and, thus, its imaginary part is an axion field with a
one-loop anomaly-induced coupling 
\begin{equation}  
{\rm one~loop~factor} \times {\rm Im}(X)
  F{\widetilde F}~. 
\end{equation} 
The supersymmetric
generalization of this coupling gives 
\begin{equation}  
{\rm ln}(X)
 W_{\alpha}W^\alpha~.
\end{equation}
Thus,  ${\rm ln}(X)$  essentially sets the value of the inverse gauge
coupling $1/g^2$. This precisely matches (\ref {SX}) and the form of 
the effective
superpotential derived from gaugino condensation in the low-energy theory for
large $X$. Indeed, for large $X$ the effective low-energy theory is pure 
Yang-Mills (plus $X$) with a  gaugino condensate. 
The effective superpotential is
simply $\langle \lambda {\lambda} \rangle \sim \Lambda^3$ where
$\Lambda$ is a low-energy scale of the theory which from the one-loop
matching of the gauge couplings at the scale $X$ is simply 
$\Lambda^3 \sim X\Lambda^3_{\rm QCD}$.
This is in agreement with the dilaton superpotential induced by
the gaugino condensation 
\begin{equation}
X \Lambda^3_{\rm QCD}\sim e^{-c/g^2} \Lambda^3_{\rm QCD}~.
\end{equation}
According to \cite {ShifmanMigdal}, the dilaton of QCD is related to 
$G_{\mu\nu}^2$ operator, which gives another way of deriving why $X$ should be 
interpreted as the field responsible for ``gluonic'' (that is, glueball) 
degrees of freedom in
accordance with  \cite {gabad} and our analysis in this paper.

\section{Discussions and Conclusions}\label{comments}

{}In this section we briefly summarize the main results of our  discussions. 
Before we turn to the conclusions we would like to 
make the following  comments. 

{}Suppose we consider now domain walls at finite $N$. Can we
use an effective superpotential (say, of the type we used here) to study them
in this case? It appears that using any type of 
truncated effective superpotential may 
not be justified. The reason why is that the gluino  condensate changes
substantially inside of the domain walls, so that integrating out heavy fields,
let us call them $Y_m$, might not be justified.
Does this imply that SQCD domain
walls are not BPS saturated at finite $N$? 
Strictly speaking one cannot draw this 
conclusion as the effective description in terms of 
$S$ and $X$ only may not be
valid, and one might have to include all of the infinitely many fields
$X,S,Y_m$ to obtain an adequate description. Practically, this means that
the problem may not be easily tractable at finite $N$.

To summarize, 
we have constructed the BPS domain walls interpolating between
neighboring vacua with broken chiral symmetry in large $N$ supersymmetric QCD.
The tension of these walls, saturating the BPS bound, scales as $\sim N$, and
the width  of the walls vanishes as $\sim 1/N$ in the large $N$ limit. 
These walls can be interpreted as supermembranes embedded in 
a (3+1)-dimensional
SQCD background. The components of the three-form supermultiplet that couples
to the supermembrane were identified in terms of composite fields of SQCD.

\acknowledgements

{}We would like to thank S. Carroll, G. Farrar, S. Lukyanov, J. Maldacena,
J. Polchinski, S. Rudaz, M. Shifman, M. Schwetz, T. Taylor, 
A. Vainshtein and  M. Voloshin 
for useful discussions. The work of G.G. was supported in part by 
the grant NSF PHY-94-23002. 
The work of Z.K. was supported in part by the grant
NSF PHY-96-02074, 
and the DOE 1994 OJI award. Z.K. would also like to thank Albert and 
Ribena Yu for financial support.

\end{document}